\documentclass[conference]{IEEEtran}
\IEEEoverridecommandlockouts
\usepackage{cite}
\usepackage{amsmath,amssymb,amsfonts}
\usepackage{algorithmic}
\usepackage{algorithm}
\usepackage{graphicx}
\usepackage{textcomp}
\usepackage{xcolor}
\usepackage{subcaption}
\def\BibTeX{{\rm B\kern-.05em{\sc i\kern-.025em b}\kern-.08em
    T\kern-.1667em\lower.7ex\hbox{E}\kern-.125emX}}
\begin{document}

\title{Symmetric Reduction Techniques for Quantum Graph Colouring\\
}

\author{\IEEEauthorblockN{Lord Sen}
\IEEEauthorblockA{\textit{Computer Science and Engineering} \\
\textit{National Institute of Technology Rourkela}\\
Rourkela, Odisha \\
123cs0226@nitrkl.ac.in}
\and
\IEEEauthorblockN{Shyamapada Mukherjee}
\IEEEauthorblockA{\textit{Computer Science and Engineering} \\
\textit{National Institute of Technology Rourkela}\\
Rourkela, Odisha \\
mukherjees@nitrkl.ac.in}
}

\maketitle

\begin{abstract}
This paper introduces an efficient quantum computing method for reducing special graphs in the context of the graph coloring problem. The special graphs considered include both symmetric and non-symmetric graphs where the axis passes through nodes only, edges only, and both together. The presented method reduces the number of coloring matrices, which is important for realization of the number of quantum states required, from $K^{N}$ to $K^{\frac{N+m}{2}}$ upon one symmetric reduction of graphs symmetric about an axis passing through $m$ nodes, where $K$ is the number of colours required and \emph{N} being total number of nodes. Similarly for other types also, the number of quantum states is reduced. The complexity in the number of qubits has been reduced by $\delta C_q= \frac{9N^2}{8}-\frac{3m^2}{8}-\frac{3Nm}{4}-\frac{N}{4}+\frac{m}{4}$ upon one symmetric reduction of graphs, symmetric about an axis passing through $m$ nodes and other types as presented in the paper. Additionally, the number of gates and number of iterations are reduced massively compared to state-of-the-art quantum algorithms. Like for a graph with 20 nodes and symmetric line passing through 2 nodes, the number of iterations decreased from 5157 to 67. Therefore, the procedure presented for solving the graph coloring problem now requires a significantly reduced number of qubits compared to before. The run time of the proposed algorithm for these special type of graphs are reduced from $O(1.9575^{N})$ to $O(1.9575^{(\frac{N+m}{2})})$ upon one symmetric reduction of graphs symmetric about an axis passing through $m$ nodes and similarly for others cases.
\end{abstract}

\begin{IEEEkeywords}
Graph coloring, Symmetry graph, Cut, Quantum Computing, Qubit, Quantum Circuits
\end{IEEEkeywords}

\begin{figure*}[ht]
\begin{subfigure}{.29\textwidth}
  \centering
  % include first image
  \includegraphics[height= 30mm,width=.80\linewidth]{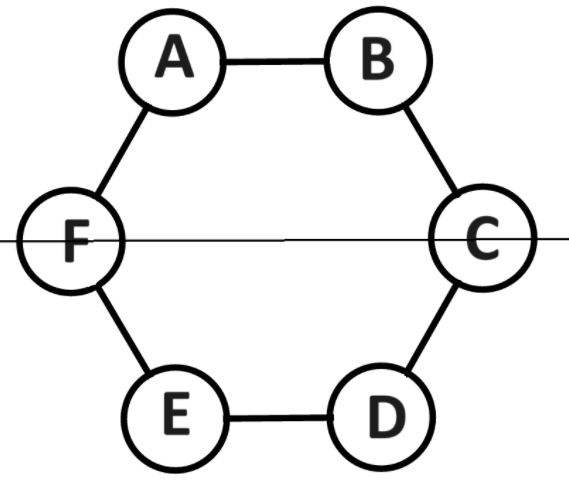}  
  \caption{}
  \label{Clustering}
\end{subfigure}
\begin{subfigure}{.29\textwidth}
  \centering
  % include first image
  \includegraphics[height= 25mm,width=.60\linewidth]{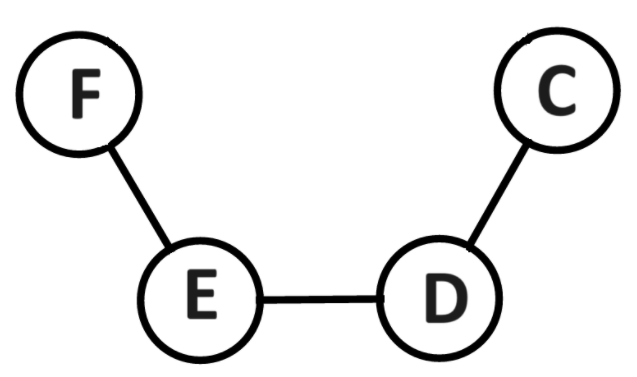}  
  \caption{}
  \label{Clustering}
\end{subfigure}
\begin{subfigure}{.29\textwidth}
  \centering
  % include first image
  \includegraphics[height= 25mm,width=.60\linewidth]{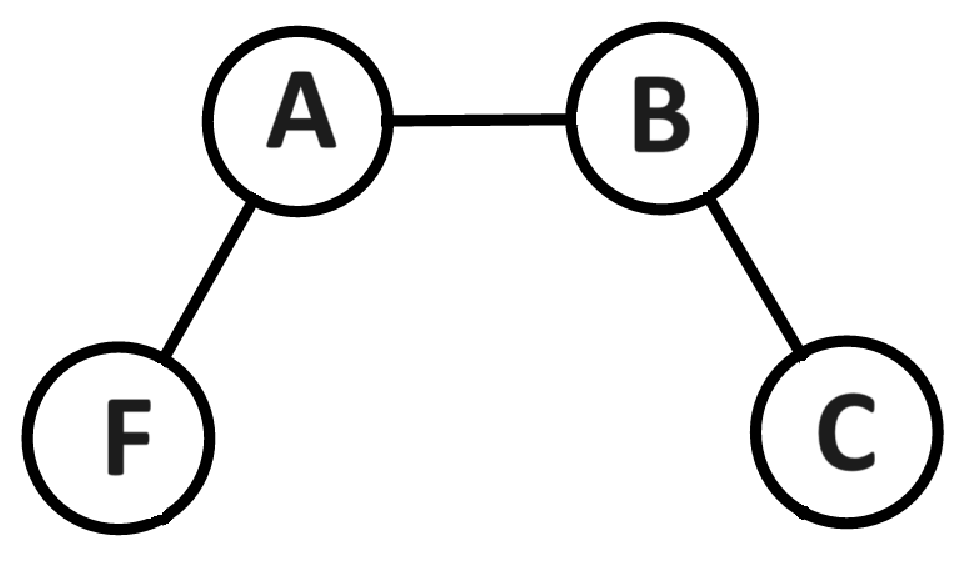}  
  \caption{}
  \label{Clustering}
\end{subfigure}
\caption{ (a) Symmetric graph with axis passing through nodes only; Cutting the graph along FC (b) Lower half (c) Upper half are obtained}
\label{g1}
\end{figure*}
\section{Introduction}
In this article, the K colouring problem of some special graphs with efficient Quantum Computing method has been presented. Classically, a \emph{K}-coloring problem for undirected graphs is the assignment of colors to the nodes of the graph such that no two adjacent nodes have the same colour, and at most \emph{K} colours are used to complete the  colouring of the graph.\par

The problem of graph colouring is age old and since then various attempts and improvements has been done. Some of them are discussed in this section. In \cite{b17}, 3-colouring problem in time $O(1.3289^n)$ has been solved. Then, an improved algorithm for 2- and 4-colouring has been presented in \cite{b18}. Thereafter, various techniques like reinforcement learning, graph neural networks have been employed for solving graph colouring problem in \cite{b19} and \cite{b20} respectively. In \cite{b21}, a solution driven multilevel approach combining coarsening, decoarsening and refinement procedure has been depicted. A new distributed graph colouring algorithm has been presented in \cite{b26}. A framework combining deep neural networks and classical heuristics have been presented in \cite{b23} and a tree based approach in \cite{b25}.\par
In \cite{b1}, Clerc has suggested a quantum based polynomial complexity algorithm to solve the \emph{K} colouring problem. In \cite{b2}, Shimizu and Mori have presented an exponential space quantum algorithm to compute the chromatic number in $O(1.9140^n)$ using QRAM and graph 20-colouring problem without using QRAM in $O(1.9575^n)$ and a method of mapping graph coloring to quantum annealing has be reported in \cite{b6},\cite{b7} and that of constrained quantum annealing in \cite{b9}. Moreover a method of graph colouring by quantum heuristics is presented in \cite{b8} and that by a reduced genetic algorithm in \cite{b3}. In addition, some discussion on symmetricity of graph colouring has been done in \cite{b4},\cite{b10},\cite{b11},\cite{b13},\cite{b14}. A colouring problem for infinite graphs has been discussed in \cite{b16} and weighted graph colouring in \cite{b22}.\\

In all the aforementioned papers the run time complexity, number of qubits, gates or number of iterations, all depend on the number of nodes of the graph. So, if the effective number of nodes, that is given as input to a quantum computer, can be reduced then the respective complexities will decrease. Our paper does the same by presenting procedures to reduce the graph.\\ 
 
In the paper, first special graphs and its types are introduced and defined with some special cases in them in Section II. IN Section III, the procedure for reduction and colouring of those are are presented using algorithms. Thereafter, the improvements achieved by the proposed reduction processes in terms of number of qubits, colouring matrices, gates, run time and number of iterations is reported and analysed via theorems and their proofs in Section IV. Lastly, the paper is summarised and future plans are suggested in the conclusion part in Section V.
\section{Special Graphs}
Here are some fundamental terminologies and symbols presented in the paper for reference.
\begin{itemize}
    \item Special graphs: The graphs coloured in this paper are refereed as special graphs for simplicity. Lets denote the set of special graphs by $\mathbf{S^*}$. 
    \item  Sub-Graphs: Parts (say $\mathbb{G_1}$, $\mathbb{G_2}$) of a graph $\mathbb{G}$ that are connected by few nodes or few edges, is called a sub-graph of $\mathbb{G}$. i.e. if $V(\mathbb{G_1}) \subset V(\mathbb{G})$ and $E(\mathbb{G_1}) \subset E(\mathbb{G}) \Rightarrow \mathbb{G_1} \subset \mathbb{G}$ then $\mathbb{G_1}$ is a sub-graph of $\mathbb{G}$.
    \item Critical Vertices: The vertices connected to the edges through which special axis of reduction passes are called critical vertices. If $i^{th}$ node is a critical node, then it is denoted by $c_i$ and its colour as $C(c_i)$. These edges through which the axis passes are called critical edges denoted by $e_{ij}$ if between nodes $c_i$ and $c_j$.
    \item Daughter graphs of order $n$ ($\mathbb{G^n}$): The graphs obtained after \emph{n} reductions of the parent graph $\mathbb{G}$ are called daughter graphs of order \emph{n} and the corresponding axes are called special axis of that order.
    
    \item Mirroring: The process of inversion of the sequence of colours of one daughter graph to obtain the colours of other symmetric daughter graph.
\end{itemize}

 \par
 In this paper special reduction methods of mostly all types of graphs are discussed, they are refered as special graphs. The graphs are classified in three types depending on the axis of section or cut:
\subsection{Case 1: Axis through nodes only:}
When the axis of section of a graph $\mathbb{G}$ passes only through the nodes of the graph then it comes under this category. This can further be classified based on the type of daughter graphs $\mathbb{G'}$ obtained or the type of axis.
\begin{itemize}
    \item \textbf{Case 1.a: Symmetric axis through nodes:}\\ Definition: An imaginary line, that passes through the nodes (only) of a graph dividing it in two parts such that both parts superimpose on each other, is called a symmetric axis through nodes and the graph is called a symmetric graph of type 1.a.\par
   The graphs which is obtained after division by the symmetric line are called daughter graphs of type 1.a. The nodes through which the line of symmetry passes is called common nodes.
   To colour the full graph, only one daughter graph needs to be coloured. The other part is obtained by mirroring it. Detailed procedure is presented in section III, Algorithm 1. \par
   Special Case: When the symmetric axis passes through two nodes only. The daughter nodes have one more than half the nodes of the parent graph. (Fig. \ref{g1}). This is a very common case; hence, discussed separately.
   \item \textbf{Case 1.b: Non-Symmetric axis through nodes:}\\ When the graph has few nodes (say $l$) in common between sub-graphs, such that a line can pass through nodes only to divide the graph into few daughter graphs then it is said to be a special graph of type 1.b.\par
Special Case:
When the graph has one node in common between two sub-graphs i.e. $l=1$, is a special case of Case 1.b as shown in Fig. \ref{Fig4}. In the figure, node 3 is the common node by which the two daughter graphs are connected. Hence, for both the daughter graphs in Fig. \ref {Fig4b} and Fig. \ref{Fig4c}, node 3 will have the same colour. Thereafter, the graphs will be coloured by the procedure as in Algorithm 3.
\end{itemize}

\subsection{Case 2: Axis through edges only}
\begin{figure}[h]
    \centering
    \includegraphics [height= 30mm,width=.65\linewidth]{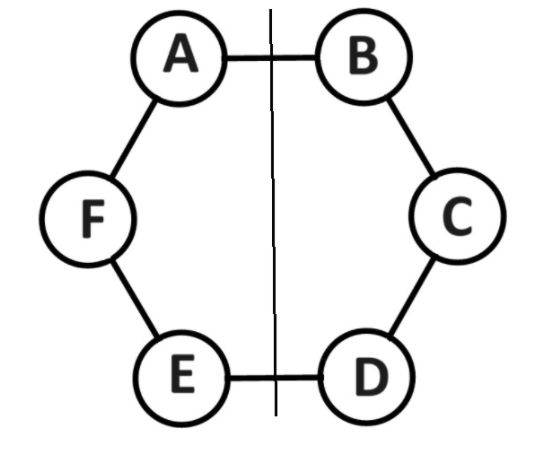}
    \caption{Graph of type 2.a: Line of symmetry passing through the edges.}
    \label{g2}
\end{figure}

When the axis of section passes only through the edges of a graph $\mathbb{G}$ then the graph falls is Case 2. It can be further classified in two types based on the nature of daughter graphs $\mathbb{G'}$. They are various classes of the case as follows:
\begin{itemize}
    \item \textbf{Case 2.a: Symmetric axis through edges}\\ An imaginary axis, that passes through the edges of a graph dividing the graph in two parts such that both parts superimpose on each other, is called a symmetric axis through edges and the graph is called a symmetric graph of type 2.a. In general the axis can pass through any number of edges. \par
    Fig. \ref{g2} shows a special case where the axis passes through two edges only. All graphs in the shape of regular polygon are one set of examples of this type.\par
    In this case both the daughter graphs are separately solved with respect to the restriction that the critical nodes have different colour. In Fig. \ref{g2}, A and F; B and G are critical nodes. The detailed procedure is in Algorithm 2.  
    \item \textbf{Case 2.b: Non-symmetric axis through edges}\\ When the graph has few edges (say $p$) between sub graphs, such that a line can pass through edges only dividing the graph in few daughter graphs then G is a special graph of type 2.b.\par
    Special Case: Two sub graphs are connected by a single edge as shown in Fig. \ref{Fig5}. Here, the graph is cut across the connecting edge. Two daughter graphs are coloured keeping colour of critical nodes different. Refer Algorithm 4 for detailed procedure.
\end{itemize}

\begin{figure}[h]
    \centering
    \includegraphics[height= 30mm,width=.80\linewidth]{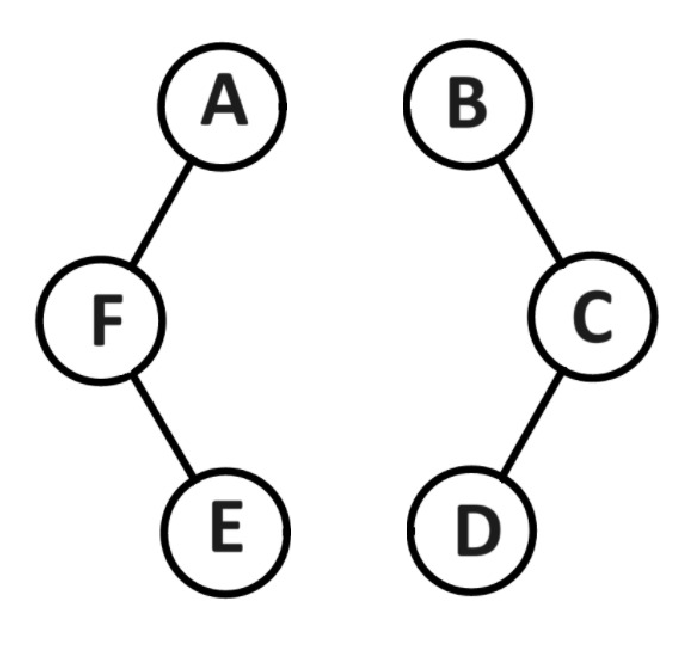}
    \caption{Two parts after division along the symmetric line of type 2.a excluding the critical edges.}
    \label{g3}
\end{figure}

\subsection{ Axis through both nodes and edges:}
Here the axis of section passes through both nodes (say $m$) and edges (say $p$) of a graph $\mathbb{G}$. This case is a combination of the first two cases as it produces the first case when $p=0$ and the second when $m=0$. Hence, this though being a more general case can be solved by combining the procedures and algorithms of the first two cases. In a similar manner this can also be further classified in two parts.
\begin{itemize}
    \item \textbf{Case 3.a: Symmetric axis through nodes and edges:}\\ 
    If an axis passes through both nodes and edges, dividing the graph in two equal parts, then the graph is a special graph of type 3.a. Algorithm 1 and 2 needs to be suitably combined, to reduce and colour this graph. \par
    Special Case 1: If the axis of symmetry passes through nodes and along edges then after section the edge along which symmetric line passes will remain. Thereafter, solving one graph and mirroring would suffice.\par
    Special Case 2: If the axis passes through a node and an edge then the parent graph will have $2n+1$ nodes and each daughter graph will have $n+1$ nodes.
    \item \textbf{Case 3.b: Non-symmetric axis through nodes and edges:}\\
    If an axis passes through both nodes and edges dividing the graph in few daughter graphs, then the axis is a special axis of type 3.b and correspondingly the graph.
\end{itemize}

\begin{figure*}[ht]
\begin{subfigure}{.29\textwidth}
  \centering
  % include first image
  \includegraphics[height= 25mm,width=.85\linewidth]{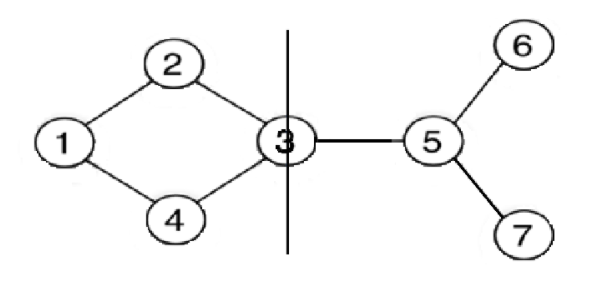}  
  \caption{ Two parts of a graph G connected by a node.}
  \label{Clustering}
\end{subfigure}
\begin{subfigure}{.29\textwidth}
  \centering
  % include first image
  \includegraphics[height= 25mm,width=.60\linewidth]{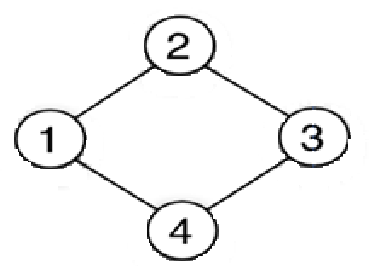}  
  \caption{ Graph G1 after cutting along the axis passing through node 3.}
  \label{Fig4b}
\end{subfigure}
\begin{subfigure}{.29\textwidth}
  \centering
  % include first image
  \includegraphics[height= 25mm,width=.60\linewidth]{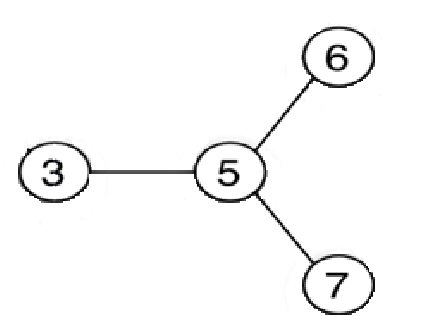}  
  \caption{Graph G2 after cutting along the axis passing through node 3.}
  \label{Fig4c}
\end{subfigure}
\caption{Non-symmetric axis passing through node.}
\label{Fig4}
\end{figure*}

\begin{figure*}[ht]
\begin{subfigure}{.29\textwidth}
  \centering
  % include first image
  \includegraphics[height= 25mm,width=.85\linewidth]{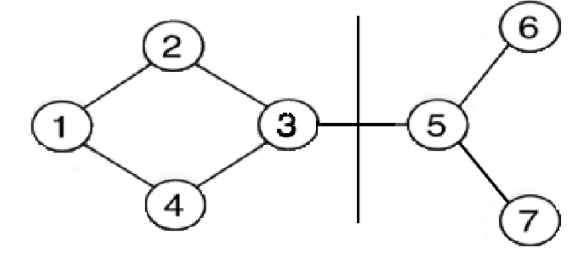}  
  \caption{Two sub-graphs connected by a critical edge $e_{35}$.}
  \label{Clustering}
\end{subfigure}
\begin{subfigure}{.29\textwidth}
  \centering
  % include first image
  \includegraphics[height= 25mm,width=.60\linewidth]{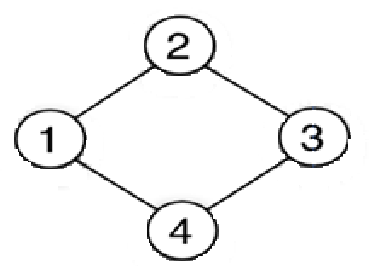}  
  \caption{ Daughter Graph G1 after cutting along the axis through $e_{35}$.}
  \label{Clustering}
\end{subfigure}
\begin{subfigure}{.29\textwidth}
  \centering
  % include first image
  \includegraphics[height= 25mm,width=.60\linewidth]{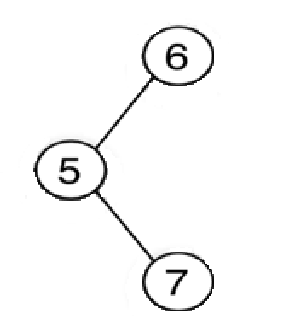}  
  \caption{Daughter Graph G2 after cutting along the axis through $e_{35}$.}
  \label{NetSelection}
\end{subfigure}
\caption{Non-symmetric axis passing through an edge connecting $c_3$ and $c_5$.}
\label{Fig5}
\end{figure*}

\section{Procedure}
In this section, the procedure of reduction and colouring of the aforementioned types of graphs has been explained in the four different algorithms 1, 2, 3, and 4. Algorithms 3 and 4 are for a single node and edge connected of type 1.b and type 2.b, respectively. Following similar algorithms, more than one node and edge connected of type 1.b and 2.b graphs can be solved. \\
After the reduction by our proposed algorithms, the reduced graphs are coloured using quantum circuits as per the quantum algorithm suggested in paper \cite{b1}, lets call it $Q$ and the sequence of colours $\mathbb{S_1}$
for a daughter graph is returned. For the process of mirroring, termination of colours of common nodes from the sequence, changing colours of critical vertices and concatenation classical and heuristic algorithms can be used. The part of reduction and colouring by quantum circuits will be mainly focused.
\begin{algorithm}[t]
Input: $~$$\mathbb{G}$ (V, E) \\
$~~~~~~~~~~$$\lvert E \rvert$ = $\mathbb{A}$\\
$~~~~~~~~~~$$\lvert V \rvert $ = $\mathbb{N}$\\
$~~~~~~~~~~$$\mathbb{K}$ = number of colours\\
Output: Sequence of colours $\mathbb{S_1}$, $\mathbb{S_2}$, $\mathbb{S}$
\caption{Symmetric Graphs Type-1.a}\label{Symmetric Graphs Type-1.a}
%\begin{multicols}{2}
\begin{algorithmic}[1]
\STATE Find the symmetric axis of $\mathbb{G}$ passing through nodes only 
\STATE Divide the graph $\mathbb{G}$ in two parts $\mathbb{G_1}$ and $\mathbb{G_2}$ along the symmetric axis\\
%\\*To solve one daughter graph $\mathbb{G_1}$\\*
\IF{$\mathbb{G_1} \in \mathbf{S^*}$}
\STATE Reduce it
\STATE Solve reduced $\mathbb{G_1}$ using $Q$
\ELSE 
\STATE Solve $\mathbb{G_1}$ using $Q$
\ENDIF
\STATE Return $\mathbb{S_1}$
\STATE Mirror $\mathbb{S_1}$ to get $\mathbb{S_2}$
\STATE Pop the corresponding colours of common nodes from $\mathbb{S_2}$ to get $\mathbb{S'_2}$
\STATE Concatenate $\mathbb{S_1}$ and $\mathbb{S'_2}$ to get $\mathbb{S}$
\STATE Return $\mathbb{S}$
\end{algorithmic}
%\end{multicols}
\end{algorithm}

\begin{figure*}[ht]
\begin{subfigure}{.19\textwidth}
    \centering
    \includegraphics[height= 25mm,width=.60\linewidth]{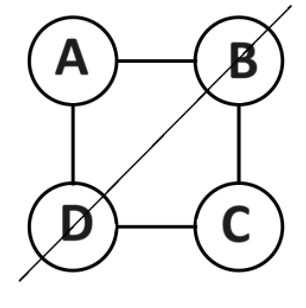}
    \caption{ Symmetric line passing through nodes only.}
    \label{figabc}
\end{subfigure}
\begin{subfigure}{.19\textwidth}
  \centering
  % include first image
  \includegraphics[height= 25mm,width=.60\linewidth]{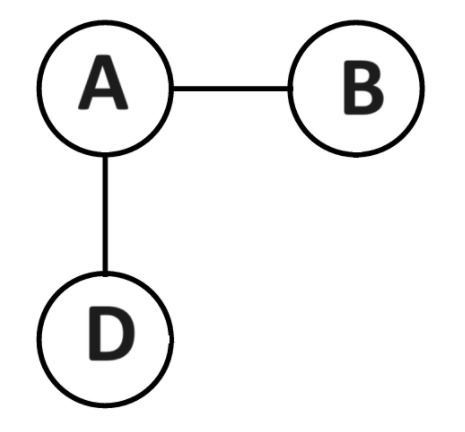}  
  \caption{Graph $\mathbb{G_1}$ after section along the symmetric axis}
  \label{fig6b}
\end{subfigure}
\begin{subfigure}{.19\textwidth}
  \centering
  % include first image
  \includegraphics[height= 25mm,width=.60\linewidth]{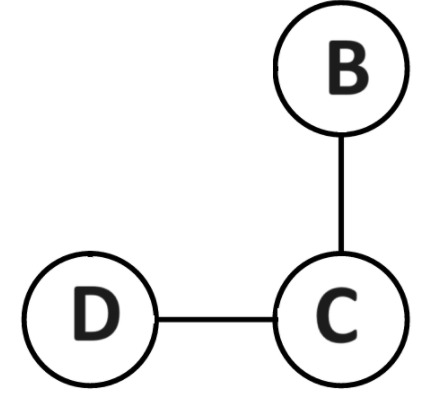}  
  \caption{Graph $\mathbb{G_2}$ after section along the symmetric axis}
  \label{fig6c}
\end{subfigure}
\begin{subfigure}{.19\textwidth}
  \centering
  % include first image
  \includegraphics[height= 25mm,width=.60\linewidth]{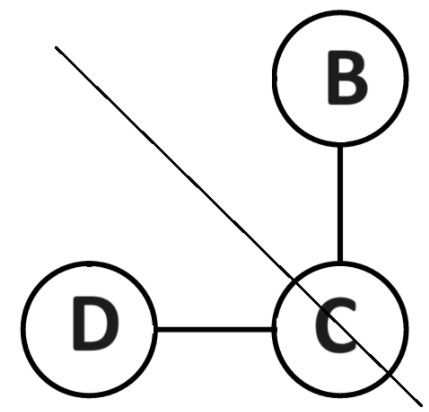}  
  \caption{Symmetric line passing through node of daughter graph $\mathbb{G_2}$}
  \label{NetSelection}
\end{subfigure}
\begin{subfigure}{.19\textwidth}
  \centering
  % include first image
  \includegraphics[height= 20mm,width=.35\linewidth]{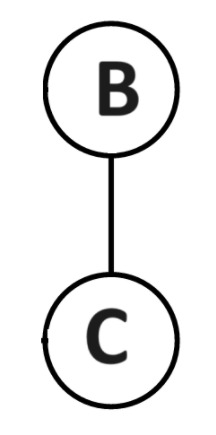}  
  \caption{Daughter graph $\mathbb{{G^2}_3}$ (of order 2) after section through the axis.}
  \label{NetSelection}
\end{subfigure}
\begin{subfigure}{.19\textwidth}
  \centering
  % include first image
  \includegraphics[height= 12mm,width=.60\linewidth]{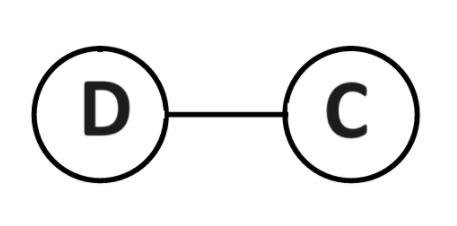}  
  \caption{Daughter graph $\mathbb{G_2}$ (of order 2) after section through the axis.}
  \label{NetSelection}
\end{subfigure}

\caption{Symmetric axis passing through the nodes of a graph G}
\label{ClusterNet}
\end{figure*}

In this part, Algorithm 1 has  been explained using a small instance of a graph depicted in Fig \ref{figabc}. The graph $\mathbb{G}$ shown in Fig \ref{figabc} contains four nodes node 1,2,3,4 represented by A, B, C, D to avoid confusion between colours and nodes. The graph $\mathbb{G}$ is sectioned along the axis of symmetry to obtain the two daughter graphs of first order namely, $\mathbb{G_1}$ with nodes B, C, D and $\mathbb{G_2}$ with nodes B, A, D as shown in Fig \ref{fig6b} and \ref{fig6c}.
Another symmetric axis through node C can be passed and the daughter graphs can be cut into further simplified version. The resulting graphs are daughter graphs of order 2. We have employed quantum circuits to solve the daughter graph $\mathbb{G_1}$.\par

Here, the graph $\mathbb{G_1}$ has $N = 3$ nodes, $A=2$ edges and $K=2$, the number of colours given(say). Hence, as per \cite{b1}, by simple permutation $K^N= 2^3= 8$ colouring matrices is required to realise all possible initial quantum states, a superposition of which is generated by the quantum circuit for initialisation. On contrast the graph $\mathbb{G}$ with nodes A, B, C, D would require  $K^N= 2^4= 16$ colouring matrices for realisation of quantum states, $K$ being the number of colours required which is 2, whereas for $\mathbb{G_3}$ with nodes B, C only 4 colouring matrices is required. The benefit of symmetric reduction is quite clear and significant from the number of colouring matrices which reduced from 16 to 8 to 4 upon two symmetric reductions. Now, each colouring matrix represents one possible colouring that may be valid or invalid. Hence, lesser the number of colouring matrices more the efficiency.\\

\begin{figure*}[ht]
\begin{subfigure}{.49\textwidth}
  \centering
  % include first image
  \includegraphics[height= 50mm,width=.70\linewidth]{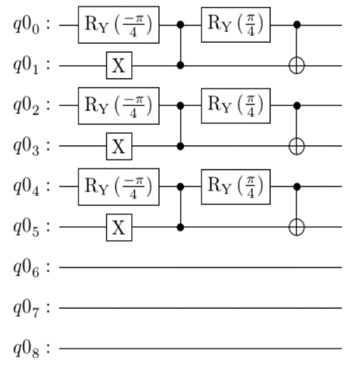}  
  \caption{Quantum Circuit for generation of colouring matrices as in \cite{b1}.}
  \label{figqc1}
\end{subfigure}
\begin{subfigure}{.49\textwidth}
  \centering
  % include first image
  \includegraphics[height= 50mm,width=0.85\linewidth]{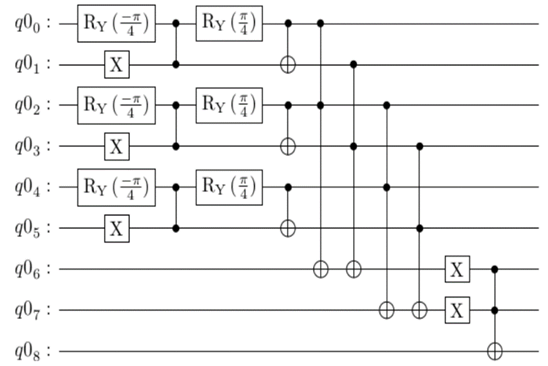}  
  \caption{Quantum Circuit for marking of valid colourings as in \cite{b1}.}
  \label{figqc2}
\end{subfigure}
\caption{Quantum Circuits for the graph colouring problem}
\label{ClusterNet}
\end{figure*}

For graph $\mathbb{G_1}$, the quantum circuit, as in \cite{b1}, for the generation of all possible quantum states is shown in Fig \ref{figqc1}. The output state is a superposition of all possible colouring sequences without consideration of the restriction that nodes connected by an edge should not have the same colour. Sequences following this restriction are valid colourings and others are invalid colourings. The valid colourings are marked by setting the ancillary qubits to $| 1>$ which are used for each edge of $\mathbb{G_1}$.The quantum circuit for the marking of valid colouring after initialization is shown in Fig \ref{figqc2}. The sequences of the superposed output state whose leftmost bit is 1 are valid colouring sequences.
Here there are 8 states, each having a probability of $\frac{1}{8}$ and are only visible in a simulation. If measured any one will be returned which may or may not be valid. Hence, there is a requirement of amplification of the valid states which is done using Grover's method \cite{b5} for amplitude amplification. As a result the quasi-probabilities of valid colourings are amplified and that of invalid are diminished. A quantum circuit for amplification can also be made following Grover's method as in \cite{b1}. For the example two amplified sequences-111011001 and 111100110 are obtained. The leftmost bit being 1 shows that these are valid colouring sequences. Removing the ancillary qubits, 1 for overall marking and 2 for marking the edges, the two solutions 011001 and 100110 are obtained, which when read from right to left are (10,01,10) and (01,10,01) which is (2,1,2) or (1,2,1).
Taking (2,1,2) for graph $\mathbb{G_1}$ and mirroring it we get the colour for $\mathbb{G_2}$ graph i.e. (2,1,2). Deleting or popping the colours of the common nodes B, D from second sequence and concatenating it with sequence of $\mathbb{G_1}$, the sequence of colours of graph $\mathbb{G}$ with nodes B, C, D, A is found i.e. (2,1,2,1) respectively.
Similarly, if the graph $\mathbb{G_1}$ would have been further reduced along C to $\mathbb{G_3}$ with nodes B, C and $\mathbb{G_4}$ with nodes C, D then, there would have been requirement of 4 quantum states only which would have further reduced the circuit in terms of number of gates, qubits and quantum states required. In that case, colour sequence for $\mathbb{G_3}$ would be (2,1) then for $\mathbb{G_4}$ would be (1,2) by mirroring. Subsequently, popping common nodes and concatenating we get sequence for $\mathbb{G_1}$ is (2,1,2) and $\mathbb{G}$ is (1,2,1,2) in sequence of A, B, C, D.

\begin{algorithm}[t]
Input: $~$Graph $\mathbb{G} = (V, E)$ \\
$~~~~~~~~~~\lvert V \rvert $ = $\mathbb{N}$\\
$~~~~~~~~~~\lvert E \rvert$ = $\mathbb{A}$\\
$~~~~~~~~~~\mathbb{K} $ = Number of colours \\
Output: Sequence of colours $\mathbb{S_1}$, $\mathbb{S_2}$, $\mathbb{S}$
\caption{Symmetric Graphs Type-2.a}\label{Symmetric Graphs Type-2.a}
%\begin{multicols}{2}
\begin{algorithmic}[1]
\STATE Find a symmetric axis of $\mathbb{G}$ passing through edges only
\STATE Divide $\mathbb{G}$ in two parts $\mathbb{G_1}$ and $\mathbb{G_2}$ along the symmetric axis
\STATE To solve $\mathbb{G_1}$, use-
\IF{$\mathbb{G_1} \in \mathbf{S^*}$}
\STATE Reduce it
\STATE Solve reduced $\mathbb{G_1}$ using $Q$
\ELSE 
\STATE Solve $\mathbb{G_1}$ using $Q$
\STATE Say $k$ colours are required
\ENDIF
\STATE Return $\mathbb{S_1}$
\STATE Colour $\mathbb{G_2}$ following steps 3 to 9 ensuring $C(c_i) \neq C(c_j)$ using these $k$ colours first.
\STATE Return $\mathbb{S_2}$
\STATE Concatenate $\mathbb{S_1}$ and $\mathbb{S_2}$ to get $\mathbb{S}$
\STATE Return $\mathbb{S}$
\end{algorithmic}
%\end{multicols}
\end{algorithm}

\begin{algorithm}[t]
Input: $~$Graph $\mathbb{G}$ = (V, E) \\
$~~~~~~~~~~$V = $\mathbb{N}$\\
$~~~~~~~~~~$E = $\mathbb{A}$\\
$~~~~~~~~~~$Number of colours $\mathbb{K}$\\
Output: Sequence of colours $\mathbb{S}$
\caption{Special Graphs Type-1.b }\label{Special Graphs Type-1.b }
%\begin{multicols}{2}
\begin{algorithmic}[1]
\STATE Find a common node of $\mathbb{G}$ connecting future $\mathbb{G_1}$ and $\mathbb{G_2}$ 
\STATE Cut $\mathbb{G}$ along the common node such that both $\mathbb{G_1}$ and $\mathbb{G_2}$ contain the common node.
\STATE Let $\mathbb{G_1}$ and $\mathbb{G_2}$ contain $n_1$ and $n_2$ nodes respectively. 
\IF{$n_1>n_2$}
\STATE Choose $\mathbb{G_2}$
\STATE To solve $\mathbb{G_2}$, use-
\IF{$\mathbb{G_2} \in \mathbf{S^*}$}
\STATE Reduce it
\STATE Solve reduced $\mathbb{G_2}$ using $Q$
\ELSE 
\STATE Solve $\mathbb{G_2}$ using $Q$
\STATE Say $k$ colours are required
\ENDIF
\ELSE
\STATE Choose $\mathbb{G_1}$
\STATE Follow steps 6 to 13
\ENDIF
\STATE Return $\mathbb{S_1}$
\STATE To solve the other daughter graph use colours from these $k$ colours first, then use any other colour if required keeping colour of common node same.
\STATE Follow steps 6 to 11 
\STATE Return $\mathbb{S_2}$
\STATE Pop the colour of the common node from $\mathbb{S_2}$ to get $\mathbb{S'_2}$.
\STATE Concatenate $\mathbb{S_1}$ and $\mathbb{S'_2}$ to get $\mathbb{S}$
\STATE Return $\mathbb{S}$
\end{algorithmic}
%\end{multicols}
\end{algorithm}

\begin{algorithm}[t]
Input: $~$Graph $\mathbb{G}$ = (V, E) \\
$~~~~~~~~~~$V = $\mathbb{N}$\\
$~~~~~~~~~~$E = $\mathbb{A}$\\
$~~~~~~~~~~$Number of colours $\mathbb{K}$\\
Output: Sequence of colours $\mathbb{S}$
\caption{Special Graphs Type-2.b}\label{Special Graphs Type-2.b}
%\begin{multicols}{2}
\begin{algorithmic}[1]
\STATE Find a common edge of $\mathbb{G}$ connecting future $\mathbb{G_1}$ and $\mathbb{G_2}$
\STATE Cut $\mathbb{G}$ along the common edge to $\mathbb{G_1}$ and $\mathbb{G_2}$
\STATE Let $\mathbb{G_1}$ and $\mathbb{G_2}$ contain $n_1$ and $n_2$ nodes respectively. 
\IF{$n_1>n_2$}
\STATE Choose $\mathbb{G_2}$
\STATE To solve $\mathbb{G_2}$, use-
\IF{$\mathbb{G_2} \in \mathbf{S^*}$}
\STATE Reduce it
\STATE Solve reduced $\mathbb{G_2}$ using $Q$
\ELSE 
\STATE Solve $\mathbb{G_2}$ using $Q$
\STATE Say $k$ colours are required
\ENDIF
\ELSE
\STATE Choose $\mathbb{G_1}$
\STATE Follow steps 6 to 13
\ENDIF 
\STATE Return $\mathbb{S_1}$
\STATE To solve the other daughter graph use colours from these $k$ colours first, then use any other colour if required, ensuring $C(c_i) \neq C(c_j)$
\STATE Follow steps 6 to 11 
\STATE Return $\mathbb{S_2}$
\STATE Concatenate $\mathbb{S_1}$ and $\mathbb{S_2}$ to get $\mathbb{S}$ 
\STATE Return $\mathbb{S}$
\end{algorithmic}
%\end{multicols}
\end{algorithm}

\section{Improvements}
The discussion in this section revolves around the advantages yielded by the proposed reduction techniques in addressing the graph coloring problem through the utilization of quantum computing. Firstly, the improvements on the basis of the quantum algorithm proposed in the paper \cite{b1} by Maurice Clerc has been reported and then the improvements in terms of run time has been compared.\par

In \cite{b1}, the number of qubits is $Q = NK + A + 1$, where Q is the number of qubits, N is the number of nodes, K is the number of colours required, A is the number of edges of the graph. In the formula $NK$ represent the number of qubits required for superposition of colouring matrix, $A$ represents the number of anciliary qubits, used to check if end colours are different or not, one for each edge and $1$ is the leftmost qubit used for denoting if its a valid state or not. Subsequently, the complexity with respect to the number of qubits is represented by $C_q$, and$C_q =\frac{3N^2}{2}-\frac{N}{2}+1$, where N is the number of nodes which can be derived by substituting N in place of K and $\frac{N(N-1)}{2}$ in place of A in the formula of Q, as $K<=N$ and maximum value of A is $\frac{N(N-1)}{2}$. The formula $Q = NK + A + 1$ is not used as the calculation done here are for general cases not a particular case. Moreover, the decrement in number of colouring matrices, which helps in realisation of quantum states, upon one reduction for each case is also showed.
\subsection{Case 1.a}
\textbf{Theorem 1:} \emph{If a graph $\mathbb{G}$ has a symmetric axis passing through only two nodes, then for each symmetric reduction, the complexity in terms of the number of qubits will decrease by $\frac{9n^2}{2}-\frac{7n}{2}-1$ where $2n$ is the number of nodes of parent graph.}\par
\textbf{Proof:}
The considered graph $\mathbb{G}$ is a symmetric graph about a line passing through two nodes. So, it must have even number of nodes, since there will be equal number of nodes on each side of symmetric axis. Let us assume, the number of nodes $N=2n$. Then the number of nodes in its daughter graph is $N'= n+1$.\par
For parent graph, $C_q=\frac{3N^2}{2}-\frac{N}{2}+1$ and   $N=2n$, Where N is the number of nodes, n being half the number of nodes and $C_q$ is the complexity in number of qubits.\par
Therefore, $C_q=\frac{3(2n)^2}{2}-\frac{2n}{2}+1=6n^2-n+1 $ where $n$ is a natural number.
For $n=2, C_q=23; n=7, C_q=288$ and so on.\par
For daughter graph $\mathbb{G'}$, $C_q'=\frac{3N^2}{2}-\frac{N}{2}+1$, where $N=n+1$ (Since, only one graph needs to be solved or considered for colouring).
\begin{center}
  $C_q'=\frac{3(n+1)^2}{2}-\frac{n+1}{2}+1 =\frac{3n^2}{2}+\frac{5n}{2}+2 $  
\end{center}

For $n=2, C_q'=13; n=7, C_q'=93$ and so on.
The difference in the complexity with respect to the number of qubits can be noticed. It is far lesser for the reduced graph with the proposed algorithm.\par
To generalise this improvement, Let 
\begin{center}
    $\delta C_q=C_q-C_q'=(6n^2-n+1)-(\frac{3n^2}{2}+\frac{5n}{2}+2)=\frac{9n^2}{2}-\frac{7n}{2}-1$
\end{center}
which is the improvement compared to the state-of-the-art method in \cite{b1}.\par
Now, 
\begin{center}
    $\frac{d(\delta C_q)}{dn}=\frac{d(\frac{9n^2}{2}-\frac{7n}{2}-1)}{dn}=9n-3.5$
\end{center}

For, $n >= 1, \frac{d(\delta C_q)}{dn} >0$. So, $\delta C_q$ is an increasing function. Hence as $n$ increases, correspondingly the number of nodes $(N=2n)$ increases, the improvement is higher.\\

\textbf{Generalised Theorem:} \emph{If the symmetric axis passes through 
$m$ nodes of a graph  $\mathbb{G}$ containing $N$ nodes then:\\
(a) The daughter graph will have $\frac{N+m}{2}$ nodes.\\
(b) If $m$ is even then $N$ is even and vice versa.\\
(c) For each symmetric reduction, the complexity in number of qubits will decrease by} ($\frac{9N^2}{8}-\frac{3m^2}{8}-\frac{3Nm}{4}-\frac{N}{4}+\frac{m}{4}$).\\

\textbf{Proof:} Let, $N$ and $m$ are the number of nodes of $\mathbb{G}$ and the number nodes the symmetric line passes through, respectively. So, remaining ($N-m$) does not contain the symmetric line. Hence, these nodes will be distributed among the two daughter graphs equally and the common nodes will remain present in both daughter graphs.\\
The number of nodes in each daughter graph is = 
$\frac{N-m}{2}+m = \frac{N+m}{2}$. Moreover, as $m<N$, So, $\frac{N+m}{2}<N$.\\
The minimum value of $m$ is 1 and in that case minimum N would be 3. So, $m>=1$ and $N>=3$ always.
For part b, it is obvious because $\frac{N+m}{2}$ being the number of nodes, is a natural number. So, if $m$ is even then $N$ must be even for $N+m$ to be divisible by 2 and vice versa.\\
Now, $C_q=\frac{3N^2}{2}-\frac{N}{2}+1$ and $C_q'=\frac{3{(N+m)}^2}{8}-\frac{N+m}{4}+1$\\
Therefore, $\delta C_q=C_q-C_q'= \frac{9N^2}{8}-\frac{3m^2}{8}-\frac{3Nm}{4}-\frac{N}{4}+\frac{m}{4}$ which is the improvement achieved.\\
Here, $C_q=\frac{3N^2}{2}-\frac{N}{2}+1$ is an increasing function for $N > 1$ and $\frac{N+m}{2}$ is less than $N$ so,  $C_q > C'_q$ and hence, $\delta C_q > 0$.
Theorem 1 is a special case of this when $m=2$ and $N=2n$.\par
Number of colouring matrices needed to represent the quantum circuit of the parent graph $\mathbb{G}$ is $K^{2n}$, where $2n$ is number of nodes of parent graph and $K$ being the number of colours required for $\mathbb{G}$, whereas for the daughter graph $\mathbb{G'}$, it is $(K')^{n+1}$, as the daughter graph will have $(n+1)$ nodes (symmetric line passes through two nodes) where $K’$ is the number of colours required for daughter graph. Clearly, as $K>1$, the number of colouring matrices decreases. Similarly for general case also the number decreases.

\subsection{Case 2.a }
\textbf{Theorem 2:} \emph{If the symmetric line passes through $m$ edges of a graph $\mathbb{G}$ containing $N$ nodes then:\\
(a) The number of nodes of $\mathbb{G}$ is even and the number of nodes of each daughter graph is $\frac{N}{2}$ which is independent of $m$.\\
(b) For each symmetric reduction, the complexity in number of qubits will decrease by $3n^2-1$.}\\

\textbf{Proof:} Whenever a symmetric line passes through an edge, it divides the corresponding nodes in two parts. So, when the graph is sectioned by the symmetric line then the total nodes is divided in two equal parts hence $\frac{N}{2}$.\par

Here $C_q$ will remain same as of Theorem 1 as the number of nodes of parent graph is same.
For each daughter graph, $C_q'=\frac{3{N'}^2}{2}-\frac{N'}{2}+1$  where $N'=n =\frac{N}{2}$ which is the number of nodes of daughter graph.\par
As both the daughter graphs needs to be coloured. So, $C_q'=2(\frac{3{N'}^2}{2}-\frac{N'}{2}+1)$
i.e. $C_q'=2(\frac{3n^2}{2}-\frac{n}{2}+1)=3n^2-n+2$
For $n=2$, $C_q'=12$, $n=7$, $C_q'=142$ and so on.
The difference can be observed in the complexity in terms of number of qubits. It is far lesser for the proposed reduced algorithm.\par 
To generalise this improvement, let 
\begin{center}
   $\delta C_q=C_q-C_q'=(6n^2-n+1)-(3n^2-n+2)=3n^2-1$ 
\end{center}
which is improvement of the proposed algorithm.\\
Now, $\frac{d(\delta C_q)}{dn}=\frac{(3n^2-1)}{dn}=6n$
For $n >= 1, \frac{d(\delta C_q)}{dn} >0.$ So, $DC_q$ is an increasing function. Hence as $n$ increases i.e. the number of qubits $(N=2n)$ increases, the improvement is higher.\par

The number of colouring matrices for the parent graph and the daughter graph are $K^{2n}$ and $2(K')^n$, respectively  where $2n$ is number of nodes of $\mathbb{G}$ and each daughter graph has $n$ nodes.

\subsection{Case 3.a }
\textbf{Theorem 3:} \emph{If the symmetric line passes through $m$ nodes and $p$ edges of a graph $\mathbb{G}$ containing $N$ nodes then:\\
a. Number of nodes of daughter graph is  $\frac{N+m}{2}$ which is independent of $p$.\\
b. If $m$ is even then $N$ is even and viceversa. \\
c. For each symmetric reduction, the complexity in number of qubits will decrease by $\frac{9N^2}{8}-\frac{3m^2}{8}-\frac{3Nm}{4}-\frac{N}{4}+\frac{m}{4}$.}\\

\textbf{Proof:} A similar proof as that of Genaralised Theorem can be done as number of nodes in daughter graph is independent of $p$ which can be shown as of Theorem 2.\\

In the next subsection of this paper a special case will be presented, when the axis of symmetry passes through one node and one edge which is $m=1$ and $p=1$. One example of this case is- the graphs in the shape of a regular polygon with odd vertices. The case is actually presented to introduce the procedure of solving this type of graphs.\\ 

\subsubsection{Special Case}
If the symmetric line passes through a node and an edge of a graph $\mathbb{G}$ containing $2n+1$ nodes then:\par
a. Number of nodes of daughter graph is  $n+1$.\par
b. For each symmetric reduction, the complexity in number of qubits will decrease by $\frac{9n^2+5n}{2}$.\\

\textbf{Proof}: As the symmetric axis passes through one node, so, remaining $2n$ nodes will be distributed in two daughter nodes. The common node will be present in both daughter graphs. So, number of nodes in daughter graphs is $\frac{2n}{2}+1 = n+1$.\\
For parent graph, $C_q=\frac{3{(2n+1)}^2}{2}-\frac{2n+1}{2}+1$ and for each daughter graph, $C'_q=\frac{3{(n+1)}^2}{2}-\frac{n+1}{2}+1$.\par
Once we have obtained the sequence of colours $\mathbb{S_1}$ of the first daughter graph $\mathbb{G_1}$ then we mirror the sequence to get another sequence.
Thereafter, the colour of the critical node needs to be suitably changed to obtain the colour sequence $\mathbb{S_2}$ for $\mathbb{G_2}$. Then the corresponding colour of common node can be terminated from $\mathbb{S_2}$ to get $\mathbb{S'_2}$. Concatenating $\mathbb{S_1}$ and $\mathbb{S'_2}$ we get the final sequence $\mathbb{S}$.\par
Hence, only one daughter graph needs to be coloured by quantum circuits.\\
Now, the improvement in the number of qubits is,\\
$\delta C_q=C_q-C_q'= (\frac{3{(2n+1)}^2}{2}-\frac{2n+1}{2}+1)-(\frac{3{(n+1)}^2}{2}-\frac{n+1}{2}+1)\\ =\frac{9n^2+5n}{2}$.\\
Clearly, $\delta C_q$ is an increasing function, since, $n>0$. So, as the number of nodes increases the improvement also increases. \\
Further, if the axis passes through $m$ nodes and one edge then the colours of m nodes needs to be terminated to get $\mathbb{S_2}$ then concatenated to get $\mathbb{S}$. Similarly, for $p$ edges and $m$ nodes, here the colours of $p$ critical nodes of each daughter graph needs to be changed.

\subsection{Case 1. b}
\textbf{Theorem 4:} \emph{If for a graph $\mathbb{G}$, there exists two sub-graphs $\mathbb{G_1}$ and $\mathbb{G_2}$ such that they are connected by a common node, then for division across that node, the complexity in number of qubits will decrease by $3(n_1-1)(n_2-1)-2$.}\\

\textbf{Proof:} Here the graph is being cut across a common node. Let the two daughter graphs have $n_1$and $n_2$ nodes respectively, then the parent graph has $N=n_1+n_2-1$ nodes. Clearly, to section about a node, at least 3 nodes are required in the parent graph. Hence, both $n_1,n_2>=2$.
For parent graph, $C_q=\frac{3N^2}{2}-\frac{N}{2}+1$,  $N=n_1+n_2-1$;
Substituting, $C_q=\frac{3(n_1+n_2-1)^2}{2}-\frac{(n_1+n_2-1)}{2}+1 =\frac{3(n_1+n_2)^2}{2}-\frac{7(n_1+n_2)}{2}+3$  \par
Let $C_{q1}$ and $C_{q2}$ be the complexity in terms of number of qubits for the daughter graphs. Then,
$C_{q1}=\frac{3n_1^2}{2}-\frac{n_1}{2}+1$ and  $C_{q2}=\frac{3n_2^2}{2}-\frac{n_2}{2}+1$\par
So, the total complexity of number of qubits is $C_q'=C_{q1}+C_{q2}=\frac{3(n_1^2+n_2^2)}{2}-\frac{(n_1+n_2)}{2}+2$
To generalise the improvement, 
let $\delta C_q=C_q-C_q'=3n_1 n_2-3n_1-3n_2+1=3(n_1-1)(n_2-1)-2$.
Clearly, since both $n_1,n_2>=2$, so, $\delta C_q>=1$ and it increases as the number of nodes increases i.e. $\delta C_q=25$ for the graph used in the example of Case 2.a.\par

The number of colouring matrices for the parent graph is $K^{n_1+n_2-1}$, where $n_1+n_2-1$ is number of nodes. For daughter graphs, number of colouring matrices are $K_1^{n_1} $  and $K_2^{n_2}$, as daughter graphs will have $n_1$ and $n_2$ nodes. As $K_1,~K_2,~n_1,~n_2>=1$ so, it can be concluded as $K^{n_1+n_2-1}>=K_1^{n_1}+K_2^{n_2}$. Hence, the number of colouring matrices are decreased by graph reduction.

\subsection{Case 2.b}
\textbf{Theorem 5:} \emph{If for a graph $\mathbb{G}$, there exists two sub-graphs $\mathbb{G_1}$ and $\mathbb{G_2}$ such that they are connected by a common edge, then for division across that edge, the complexity in number of qubits will decrease by $3n_1 n_2-1$.}\\

\textbf{Proof:} The parent graph is cut across an edge. Let the two daughter graphs have $n_1$and $n_2$ nodes respectively, then the parent graph has $N=n_1+n_2$ nodes.\\
For parent graph, $C_q=\frac{3(n_1+n_2)^2}{2}-\frac{(n_1+n_2)}{2}+1$, $N=n_1+n_2$\\
For daughter graphs, $C_{q1}=\frac{3n_1^2}{2}-\frac{n_1}{2}+1$ and  $C_{q2}=\frac{3n_2^2}{2}-\frac{n_2}{2}+1$\\
Therefore, $\delta C_q=C_q-C_q'=3n_1 n_2-1$\\
As, $n_1 ,n_2>=1$ so, $\delta C_q>=2$ and it increases as the number of nodes increases.
$\delta C_q=35$ for the graph used in the example of Case 2.b. where  $n_1=4$, $n_2=3$.\par
The number of colouring matrices for the parent graph is $K^{n_1+n_2 }$, where $n_1+n_2$ is number of nodes.\par
For daughter graphs it is $K_1^{n_1}$  and $K_2^{n_2}$, as daughter graphs will have $n_1$ and $n_2$ nodes. As $K_1,~K_2,~n_1,~n_2>=1$ so, we can say that $K^{n_1+n_2}>=K_1^{n_1} +K_2^{n_2}$. Hence, the number of colouring matrices decreases by graph reduction.
Similarly, for all the cases we can show that the number decreases.

\subsection{Comparison of Cases:}
\textbf{Theorem 6:} \emph{If a graph $\mathbb{G}$ falls both in type 1.a and 2.a then-\\
a. It's better to opt type 2.a if $m>= \frac{3N}{7}$.\\
b. It's better to opt type 1.a if $m< \frac{N}{2.4446303}$.}\\

\textbf{Proof:} For symmetric graph of type 1.a having $N$ nodes and the symmetric line passes through $m$ nodes then the improvement in the number of qubits is  $\delta C_{q1} = \frac{9N^2}{8}-\frac{3m^2}{8}-\frac{3Nm}{4}-\frac{N}{4}+\frac{m}{4}$ where $N >=3, m>=1, m<N$ and $m,N$ are Natural numbers.\\
For symmetric graph of type 2.a having $N$ nodes and symmetric axis passes through $p$ edges then the improvement is $\delta C_{q2}= \frac{3N^2}{4}-1$ where $N>=2$,$p>=1$, $N=2n$ where $p,n$ are natural numbers.\\
 Now, if a graph falls in both the category then $N>=4$, $m>=2$, $p>=1$, $m<N$, $N=2n$ and $m,N,p$ are natural numbers.\\
 Let, $\delta= \delta C_{q1}-\delta C_{q2}$. So, if $\delta > 0$ then $\delta C_{q1}> \delta C_{q2}$, so, the profit in solving by the technique of type 1.a is more and if $\delta < 0$ then $\delta C_{q1}< \delta C_{q2}$, so the profit in solving by 2.a is more.\\
 Here, $\delta =\delta C_{q1}-\delta C_{q2}= \frac{3N^2}{8}-\frac{3m^2}{8}-\frac{3Nm}{4}-\frac{N}{4}+\frac{m}{4}+1$.\\

\begin{figure*}[ht]
 \begin{subfigure}{.49\textwidth}
  \centering
  % include first image
  \includegraphics[height= 45mm,width=.80\linewidth]{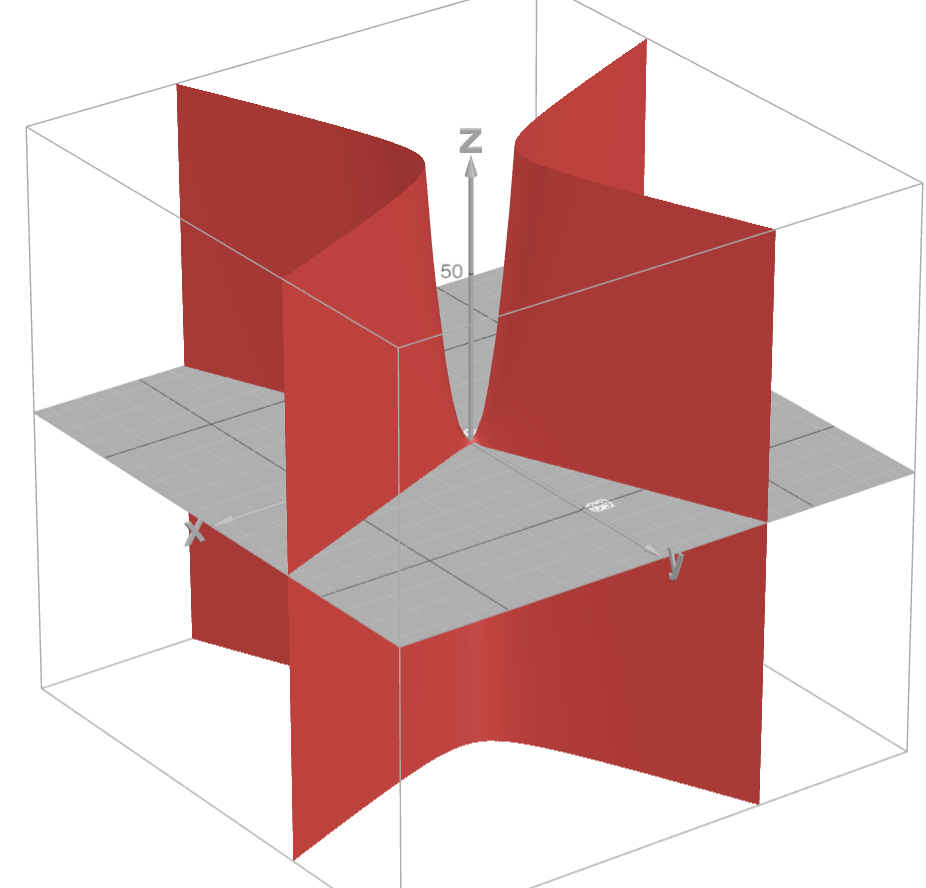}  
  \caption{ The graph of $\delta (x,y)$ without any restriction.}
  \label{figdes3}
\end{subfigure}
 \begin{subfigure}{.49\textwidth}
  \centering
  % include first image
  \includegraphics[height= 45mm,width=.80\linewidth]{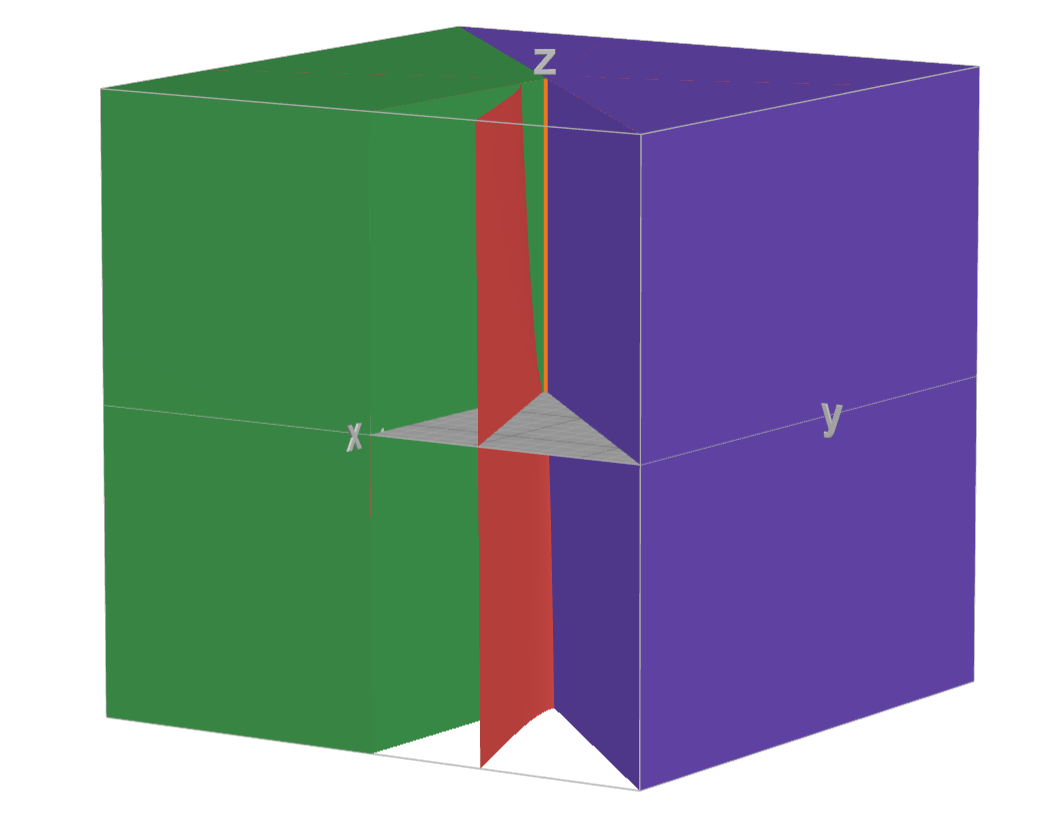}  
  \caption{$\delta (x,y)$ under the restriction that $x>=4$, $y>=2$, $y<x$. The graph is shown with red color. It is clear that $\delta (x,y)$ takes both positive and negative values. }
  \label{figdes4}
\end{subfigure}
\begin{subfigure}{.49\textwidth}
  \centering
  % include first image
  \includegraphics[height= 45mm,width=.80\linewidth]{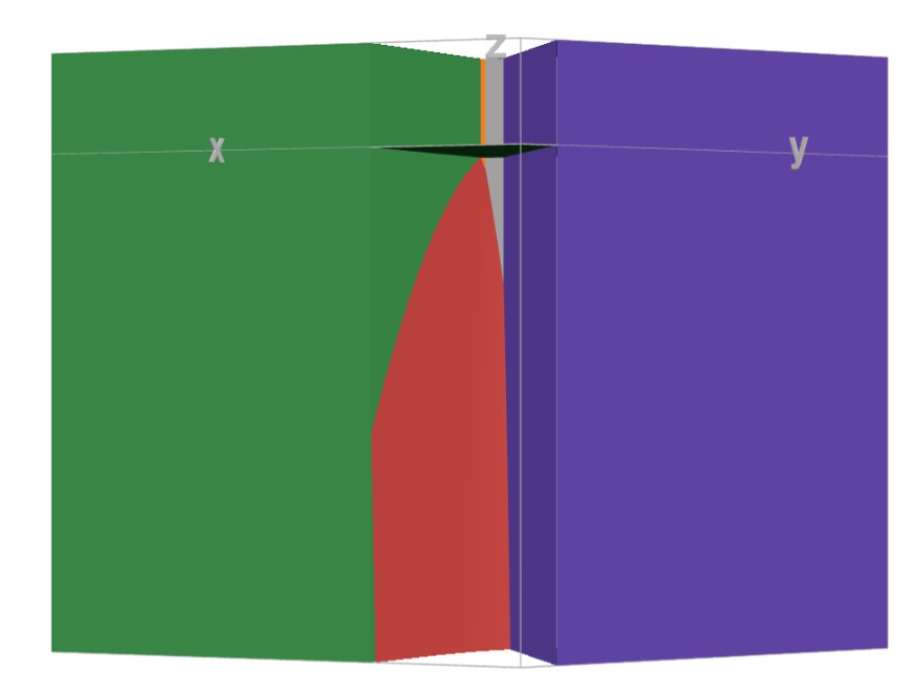}  
  \caption{$\delta (x,y)$ under the restriction that $x>=4$, $y>=2$, $y<x$ and one additional restriction that $y>=\frac{3x}{7}$. Clearly, $\delta (x,y) <0$ for all values of $x$ and $y$ following this restriction.}
  \label{figdes1}
\end{subfigure}
\begin{subfigure}{.49\textwidth}
  \centering
  % include first image
  \includegraphics[height= 45mm,width=.80\linewidth]{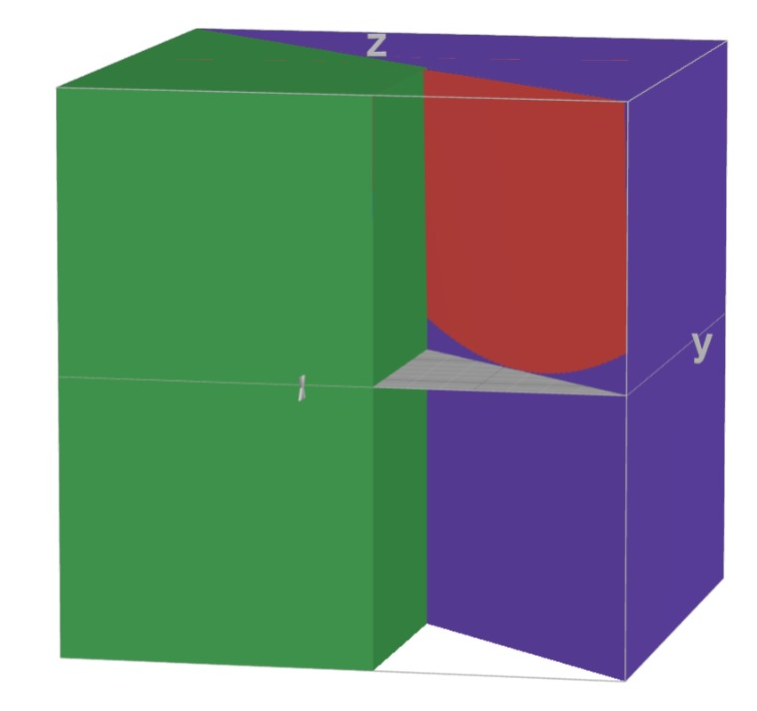}  
  \caption{$\delta (x,y)$ under the restriction that $x>=4$, $y>=2$, $y<x$ and one additional restriction that $y<=\frac{x}{2.4446303}$. Clearly, $\delta (x,y) >0$ for all values of $x$ and $y$ following this restriction.}
  \label{figdes2}
\end{subfigure}
\caption{ $\delta (x,y)$ under certain conditions to find its sign that help in realisation of which technique to be used. }
\label{ClusterNet}
\end{figure*}

 There is a requirement to find the interval where $\delta > 0$ and $\delta < 0$. Replacing N and m by x and y we have, $\delta (x,y)= \frac{3x^2}{8}-\frac{3y^2}{8}-\frac{3xy}{4}-\frac{x}{4}+\frac{y}{4}+1$, Fig \ref{figdes3} where $x>=4$, $y>=2$, $y<x$.
 Hence, the region where $x<4$, $y<2$ and $y>x$ is not our region of interest. Hence, we are blocking it in Fig \ref{figdes4}.\\
 Analysing the graph of $\delta (x,y)$ in the restricted region, we get when $y>=\frac{3x}{7}$ then $\delta (x,y) <0$ as shown in Fig \ref{figdes1} and for $y<\frac{x}{2.4446303} $ we have $\delta (x,y) >0$ as shown in Fig \ref{figdes2}. \\
 So, for $m>=\frac{3N}{7}$, $\delta C_{q1}< \delta C_{q2}$. Hence, its favourable to solve by technique employed in graph of type 2.a. 
 For $m<\frac{N}{2.4446303} $, $\delta C_{q1}> \delta C_{q2}$. Hence, its favourable to solve by technique employed in graph of type 1.a.\\
 
But, when $\frac{x}{2.4446303}<y<\frac{3x}{7}$ then the nature of the graph is ambiguous and it changes sign very steeply. In this case the value of $m$ and $N$ needs to be substituted in $\delta$ to find its sign and decide the procedure accordingly. Moreover the interval $(\frac{N}{2.4446303},\frac{3N}{7})$ is so small that for an natural number $m$ to lie there, the value of $N$ must be greater than or equal to $50$.\\

In a similar manner comparison can be done if a graph falls in multiple categories. This way the optimal procedure to reduce a graph can be found.

\subsection{Analysis of number of Gates:}
As per \cite{b1}, the maximum number of quantum gates required is $4N^2-3N$ for initialisation, $\frac{N^3-N^2}{2}$ for marking of valid colouring and $N^3+11N^2-8N+9$ for amplification, considering $K=N$ and number of edges $A= \frac{N(N-1)}{2}$. So, total is $\frac{3N^3}{2}+ \frac{29N^2}{2}- 11N + 9$. This is an increasing function.\\
Using the reduction techniques presented in the paper the number of quantum gates required can be significantly reduced. This can be easily proved for each case by replacing $N$ by the number of nodes in daughter graph and then following similar procedure as done for number of qubits. To visualise the massiveness of the reduction, let us consider an example of a symmetric graph of type 1.a with 10 nodes and symmetric line through 2 nodes, which will require 2849 gates. After reduction it will have 6 nodes hence, the number of gates reduced to 789. Similarly, the overall size of the circuit also decreases and it decreases even more rapidly as the overall complexity is $O(N^4)$ in worst case.

\subsection{Analysis of number of Iterations:}
After the generation of valid colourings, their amplification needs to be done. The amplification is done by Grover's method. If there are $P$ possible states with 1 valid solution then, the number of iterations required to amplify the valid colourings is $\frac{\pi}{4}\sqrt{P}$. Now, if there are $k$ valid solutions then the number of iterations becomes $\frac{\pi}{4}\sqrt{\frac{P}{k}}$.

As, per \cite{b1}, the number of states $P$ is $K^N$ and $k$ is of the order $N!$. Moreover, $K<= N$, So, the maximum number of iterations is $\frac{\pi}{4}\sqrt{\frac{N^N}{N!}}$ where $N$ is the number of nodes of the graph.
The proposed algorithms in this paper reduce the number of iterations massively as the number of nodes of the graph to be solved by a quantum circuit decreases. For example, if there exits a symmetric graph of type 1.a with 20 nodes. Then after one symmetric reduction by a line through 2 nodes the daughter graph will have 11 nodes. Now, the number of iterations for parent graph is $\frac{\pi}{4}\sqrt{\frac{20^{20}}{20!}}= 5157$ iterations. On contrary, the number of iterations required in daughter graph is $\frac{\pi}{4}\sqrt{\frac{11^{11}}{11!}}= 67$ iterations only. This massive reduction in number of iterations is one of the key achievements of this paper. In a similar way for other types of reductions also the number of iterations will decrease significantly.
\subsection{Run-time analysis:}
In this section, the improvements in with respect to run time of the proposed graph reduction techniques has been shown. For this comparison, state-of-the-art quantum algorithm proposed in  \cite{b2}. In the paper, two runtimes have been presented. One by using QRAM for Chromatic number calculation and other not using QRAM for 20 colouring problem i.e. a K colouring problem with $K<=20$ with restriction that . The corresponding running time are $O (1.9140^n)$ and $O (1.9575^n)$ respectively.
\subsubsection{Case 1.a}
Clearly as the value of number of nodes decreases from $2n$ to $n+1$ when the symmetric axis passes through two nodes. So, the running time for the K colouring problem also decreases from $O (1.9575^{2n})$ to $O (1.9575^{n+1})$ and from $O (1.9575^{N})$ to $O (1.9575^{\frac{N+m}{2}})$ fro the generalised case when the symmetric axis of type 1.a passes through $m$ nodes.\par
\subsubsection{Case 2.a}
For case 2.a when the symmetric axis passes through $p$ edges, then the run time decreases from $O (1.9575^{2n})$ to $2*O (1.9575^{n})$ as the number of nodes decreases from 2n to n and the colouring of both daughter graphs need to be done.\par

Similarly, for the remaining cases the running time will decrease as there is reduction in the effective number of nodes in the graph to be passed to the quantum circuit.

\section{Conclusion}
 In this paper optimal graph reduction techniques of some special graphs was presented for obtaining an efficient quantum based method for solving the K colouring problem of these graphs.\par
The proposed reduction techniques have helped achieve improvements, by reducing the number of colouring matrices used in realisation of quantum states, complexity in terms of number of qubits and the number of quantum gates. The running time and the number of iterations has been decreased significantly by the proposed algorithm. \\
In future, the algorithm 2 and correspondingly Theorem 2 can further optimized. There instead of solving both the daughter graphs, we can solve one then mirror the sequence to get other and correspondingly change the colour of critical nodes or another algorithm can be developed that will just mirror the non critical vertices and suitably colour the critical vertices following the restriction of graph colouring.

\vspace{12pt}

\end{document}